\documentclass[%
 reprint,
 amsmath,amssymb,
 aps,
 prl,
]{revtex4-1}

\usepackage{graphicx}
\usepackage{dcolumn}
\usepackage{bm}

\usepackage{todonotes}

\newcommand{\ie}{{\it i.e. }}

\begin{document}

\preprint{APS/123-QED}

\title{Rigidity Transition of Jammed Packings changes \\ Universality Class at Weak Attraction.}
\title{Sticky Matter:\\Jamming and rigid cluster statistics with attractive particle interactions}

\author{Dion J. Koeze}
	\email{d.j.koeze@tudelft.nl}
\author{Brian P. Tighe}
\affiliation{
 Delft University of Technology, Process \& Energy Laboratory, Leeghwaterstraat 39, 2628 CB Delft, The Netherlands.
}

\date{\today}

\begin{abstract}
While the large majority of theoretical and numerical studies of the jamming transition consider athermal packings of purely repulsive spheres, real complex fluids and soft solids generically display attraction between particles. By studying the statistics of rigid clusters in simulations of soft particles with an attractive shell, we present evidence for two distinct jamming scenarios. Strongly attractive systems undergo a continuous transition in which rigid clusters grow and ultimately diverge in size at a critical packing fraction. Purely repulsive and weakly attractive systems jam via a first order transition, with no growing cluster size. We further show that the weakly attractive scenario is a finite size effect, so that for any nonzero attraction strength, a sufficiently large system will fall in the strongly attractive universality class. We therefore expect attractive jamming to be generic in the laboratory and in nature.
\end{abstract}

\maketitle

\begin{figure}[b]
	\centering
    \includegraphics[width=\columnwidth]{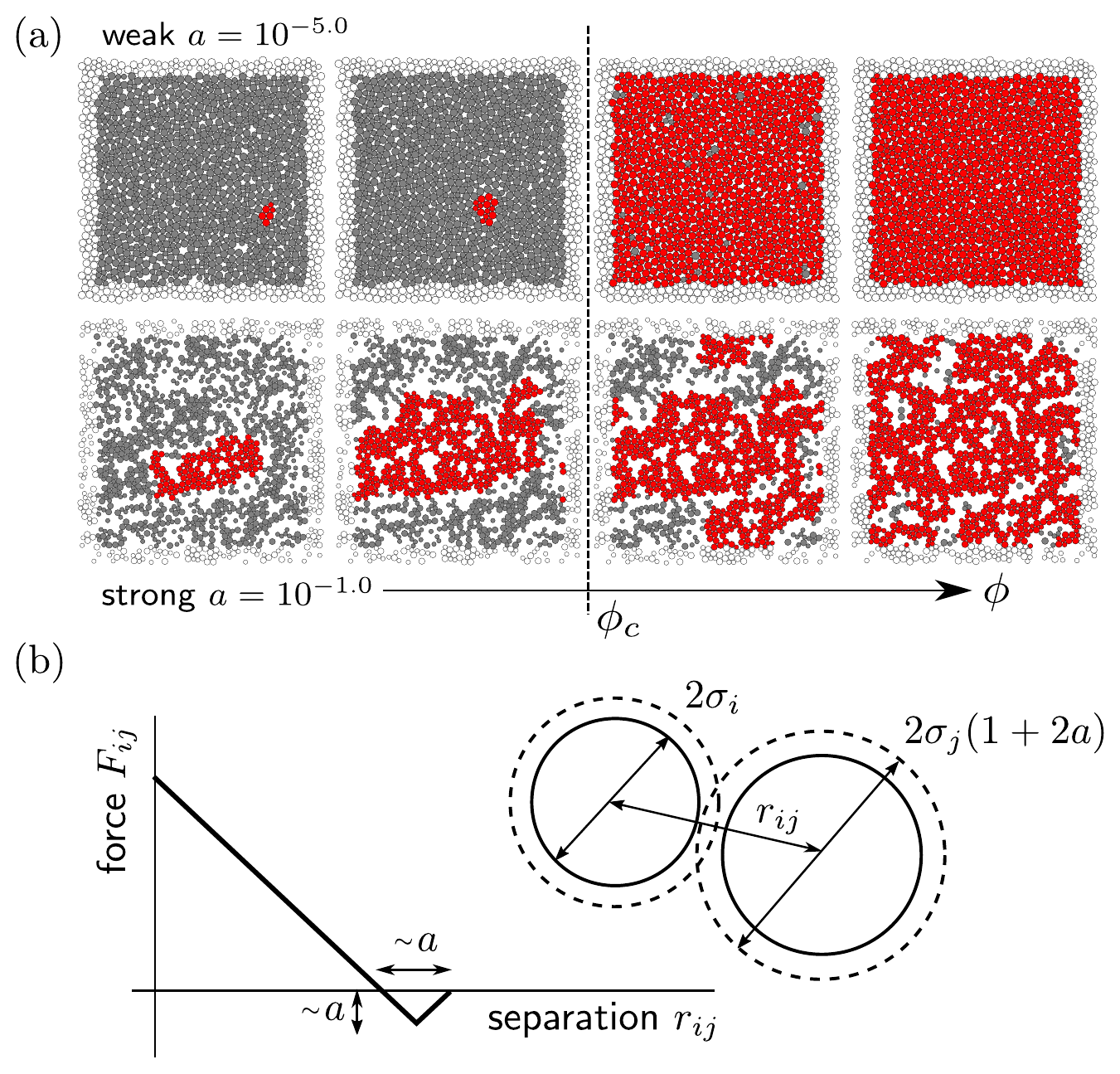}
	\caption{(a) Packings with weak and strong attraction for packing fractions $\phi$ near the point $\phi_c$ where they jam. Particles in red form the largest rigid cluster. (b) Contact force law for a pair of particles with an attractive shell.}
	\label{fig:cartoon}
\end{figure}

Numerous complex fluids, including emulsions, foams, pastes, powders, sand, and blood, can jam into soft amorphous solids under increasing packing fraction \cite{vanhecke10,liu10a}. In recent years, enormous progress towards a fundamental understanding of jammed matter has been driven by theoretical and numerical studies of dense systems of athermal spheres interacting via purely repulsive contact forces. There is now general agreement on how the structure and mechanics of repulsive soft spheres are governed by proximity to the jamming transition at a critical packing fraction $\phi_c$ -- see e.g.~\cite{durian95,ohern03,wyart05b,wyartannales,ellenbroek09,olsson07,ikeda12,mizuno16b,boschan16,baumgarten17b} for a partial list. This line of study implicitly builds on the assumption that repulsive particles yield broad or even universal insights into the marginally jammed state. Nevertheless, purely repulsive interactions are not generic in the laboratory or in nature. While stickiness has various origins (e.g.~van der Waals forces \cite{hutzlerweairebook}, depletion effects \cite{becu06,jorjadze11}, wetting effects \cite{herminghaus05,moller07,yunker11}, interface deformation \cite{vella05,karpitschka16}, critical Casimir forces \cite{bonn09}, etc.), particles typically attract their neighbors, and pure repulsion can only be realized with careful tuning, if at all. 
The few existing studies of jamming with attraction reveal significant differences, including a gel-like structure with large voids \cite{head07,zheng16} and shear banding \cite{chaudhuri12,irani14,irani16,yamaguchi17}. Most remarkably, Lois et al.~\cite{lois08} showed that strongly attractive soft spheres belong to a new universality class, distinct from both repulsive jamming and rigidity percolation on generic lattices \cite{ellenbroek15,henkes16}. But it remains unclear when repulsive jamming gives way to attractive jamming -- one cannot currently predict whether a given experimental system falls into the repulsive or attractive jamming class.

In this Letter, we demonstrate the striking influence of attraction on the growth of rigid clusters, illustrated in Fig.~\ref{fig:cartoon}. A cluster is rigid if, when removed from the packing, its only zero frequency vibrational modes are rigid body motions. A system is jammed if it contains a spanning rigid cluster \cite{jacobs95}.
Fig.~\ref{fig:cartoon}a depicts disk packings with ``weak'' (top row) and ``strong'' (bottom row) attraction; they differ in the thickness of an attractive shell (panel b). The largest rigid cluster in each packing is shaded red. 
For weak attraction, the largest cluster contains just a few particles, and a spanning cluster appears suddenly at $\phi_c$. This scenario resembles the first order transition observed in repulsive systems \cite{ellenbroek15,henkes16}, suggesting attraction acts as a small perturbation. 
In sharp contrast, clusters in strongly attractive systems grow in size before spanning at $\phi_c$, reminiscent of a continuous phase transition with a diverging length scale. 

What distinguishes repulsive, weakly attractive, and strongly attractive jamming?
Here we use rigid cluster decomposition to identify the attractive jamming point and to quantitatively assess the order of the jamming transition. Then, by systematically varying attraction and particle number, we determine when weakly attractive jamming ends and strongly attractive jamming begins. Our central result is that attraction is never weak in the limit of asymptotically large system sizes -- large systems are either purely repulsive or strongly attractive, and any amount of attraction places a system in the universality class of strongly attractive jamming.

{\em Methods and protocol.---} We consider athermal systems of $N$ disks in a 50:50 bidisperse mixture with size ratio 1.4:1 to avoid crystallization \cite{ohern03,koeze16} and periodic boundary conditions to eliminate wall effects. Unless stated otherwise we choose $N=1024$. Athermal attractive particles are strongly protocol-dependent, because contacts can only break or form through external excitation. We employ a standard preparation protocol in which particles are initially placed at random, followed by a quench at fixed $\phi$ to a local energy minimum using a nonlinear conjugate gradient method \cite{ohern03}. Note that, unlike repulsive jamming, the the jamming point cannot be identified with zero pressure, as tensile states are accessible \cite{head07}.

We adopt the conventions of prior work \cite{lois08,chaudhuri12,irani14,irani16,zheng16,yamaguchi17} and model sticky particles with a repulsive core and attractive shell that experience a central force
\begin{equation}
F_{ij} = \left\{
\begin{aligned}
& k \delta_{ij}                 &                 \delta_{ij} &\geq -\sigma_{ij} a  \\
& -k(\delta_{ij} + 2a\sigma_{ij}) & -\sigma_{ij}a > \delta_{ij} &\geq -2\sigma_{ij} a \\
& 0                              &                 \delta_{ij} &< -2 \sigma_{ij} a 
\end{aligned}
\right.
\end{equation}
between particles $i$ and $j$ (see Fig.~\ref{fig:cartoon}b).
The spring constant $k$ characterizes repulsion, while the dimensionless attraction strength $a$ sets the attractive shell thickness and the maximal tensile force. 
$\delta_{ij} = \sigma_{ij} - r_{ij}$ is the overlap between  two particles and $\sigma_{ij}$ is the sum of the radii of their cores. 
The packing fraction $\phi$ is calculated from the particles' cores. Including the attractive shell would increase $\phi$ by a factor $1 + 4a$, to leading order in $a$.

The pebble game algorithm \cite{jacobs95} efficiently and unambiguously identifies all rigid clusters in two spatial dimensions, dictating our choice to simulate disk packings. The algorithm outputs disjoint sets of bonds (i.e.~clusters) whose bonds are rigid with respect to each other. Details are found in Ref.~\cite{jacobs95}. Accurate contact identification is essential for rigid cluster decomposition. Unlike repulsive particles, identifying contacts with attraction is straightforward because particles tend to sit near the first zero of $F_{ij}$ (i.e.~the minimum of their pair potential).

\begin{figure}[tb]
	\centering
	\includegraphics[width=\columnwidth]{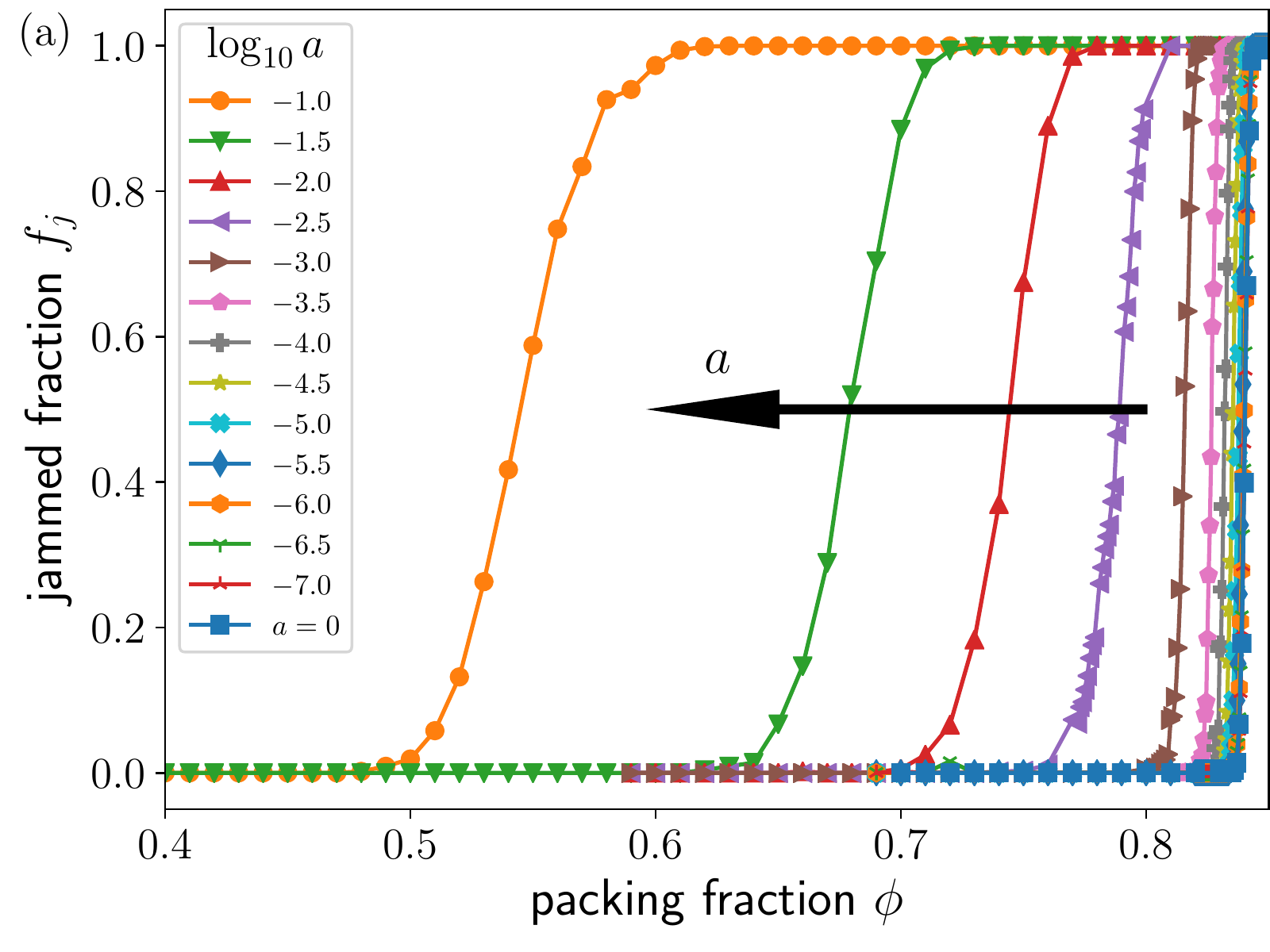}
	\includegraphics[width=\columnwidth]{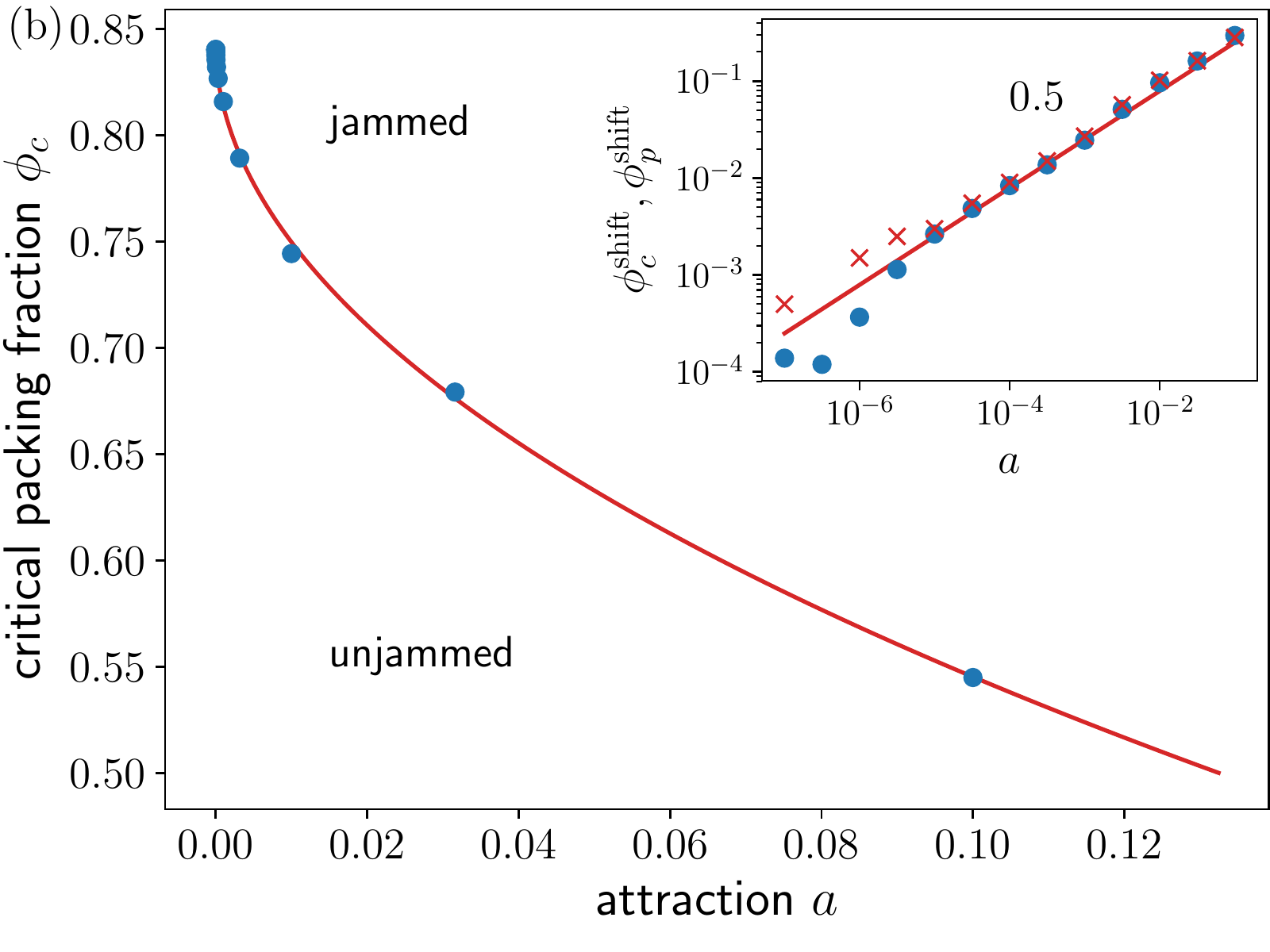}
	\caption{(a) Fraction of jammed states $f_j$ versus packing fraction $\phi$ for varying attraction strength $a$ and $N=1024$. 
(b) Attractive jamming phase diagram. 
(inset) Scaling of the shift in $\phi_c(a)$ (filled circles) and $\phi_p(a)$ (crosses) with $a$.}
	\label{fig:fj}
\end{figure}

{\it Jamming phase diagram.}--- As we are considering physics near jamming, we first determine the critical packing fraction $\phi_c$ as a function of attraction strength. 

For finite particle number $N$, the jamming transition is ``blurred'' by finite size effects, as seen in a plot of the fraction $f_j$ of jammed packings in ensembles prepared at a given $\phi$ (Fig.\,\ref{fig:fj}a). The purely repulsive packings show a rapid increase of $f_j$ at a packing fraction near $0.84$. As attraction strength $a$ increases, the rise in $f_j$ shifts to lower $\phi$ and also becomes more gradual. We will first focus on the shift and then on the widening of $f_j$.

We associate a critical packing fraction $\phi_c(a,N)$ with the value of $\phi$ where $f_j(\phi,a,N) = \Delta$ with $\Delta = 0.5$. The shift of the transition is then defined with respect to the purely repulsive jamming point, \ie $\phi_c^\mathrm{shift}(a,N) = \phi_c(0,N) - \phi_c(a,N)$. 
Henceforth we drop the $N$ dependence of $\phi_c(a)$ whenever $N=1024$. We have verified that the scaling of $\phi_c^\mathrm{shift}$ is insensitive to variations in $\Delta$ around $0.5$.
In Fig.\,\ref{fig:fj}b we see how $\phi_c(a)$ decreases with increasing $a$, dividing the diagram into unjammed and  jammed phases. 
The shift is plotted in the inset of Fig.\,\ref{fig:fj}b (filled circles). We find power law scaling that is well described by $\phi_c^\mathrm{shift} \sim a^{0.5}$. Note that the excess volume occupied by attractive shells, which scales linearly in $a$, cannot trivially account for this rapid decrease.

\begin{figure}
	\centering
	\includegraphics[width=\columnwidth]{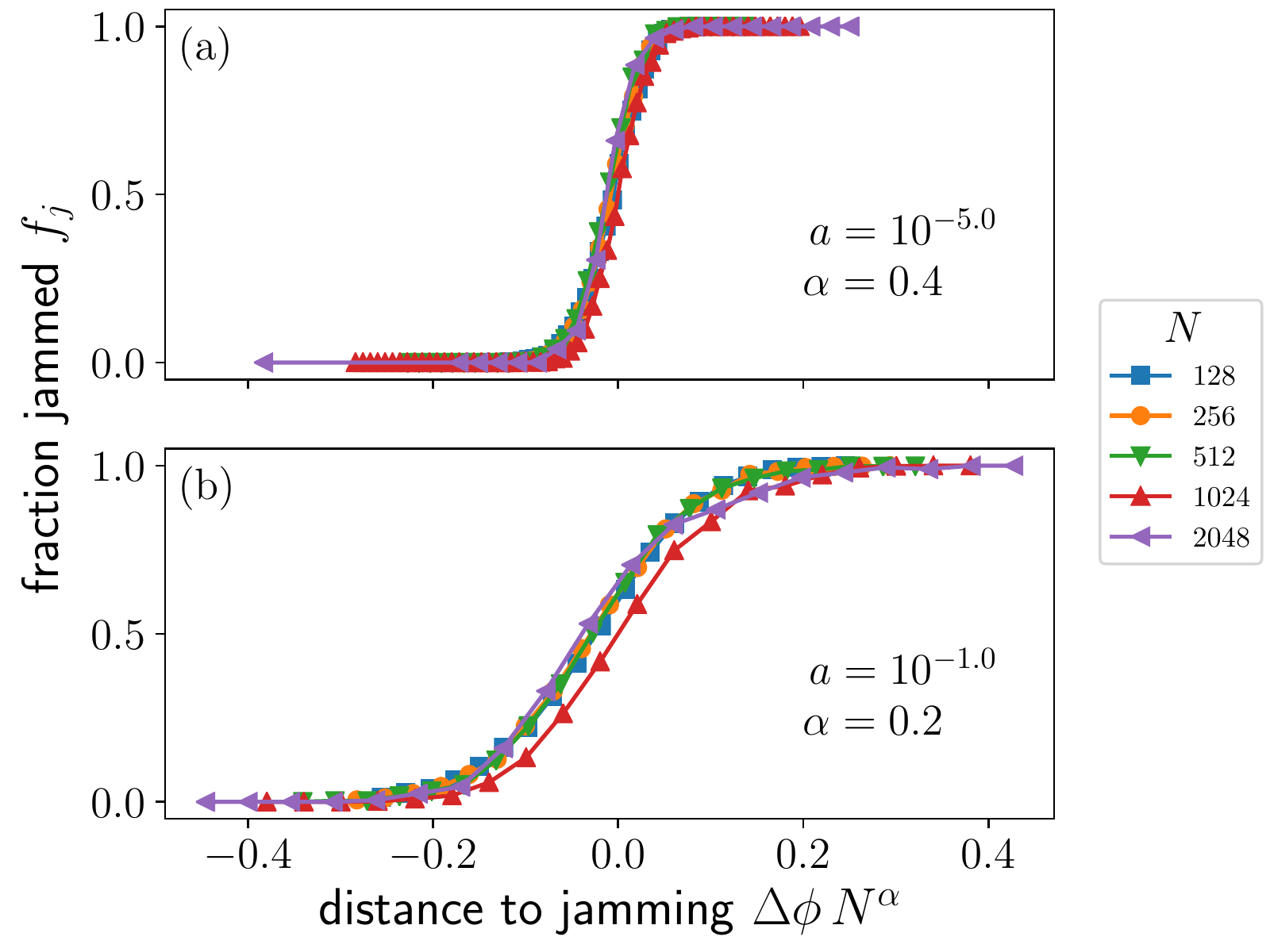}
	\caption{Data collapse of the fraction of jammed states for varying particle number $N$ in (a) weakly attractive and (b) strongly attractive systems. 
}
	\label{fig:finsize}
\end{figure}

We now ask if the jamming transition is sharp in the large system size limit. We focus on ``weak'' ($a=10^{-5.0}$) and ``strong'' ($a=10^{-1.0}$) attraction, plotted in Fig.\,\ref{fig:finsize}a and b, respectively. For $N = 128 \ldots 2048$, $f_j$ can be collapsed by plotting versus $\Delta \phi \, N^\alpha$, where $\Delta \phi = \phi - \phi_c(a,N)$. We observe data collapse for positive values of the exponent $\alpha$, hence $f_j$ approaches a step function as $N \rightarrow \infty$ and the transition is indeed sharp. However, the value of $\alpha$ providing the best collapse for the plotted range of $N$ is different for weak and strong  attraction -- $\alpha \approx 0.4$ versus $0.2$, respectively. This is the first indication in our data of a distinction between weak and strong attraction.

{\it Order of the transition.}--- The growth of rigid clusters illustrated in Fig.~1 suggests that jamming is a continuous transition in strongly attractive systems, and a first order transition in weakly attractive (or purely repulsive) systems. 
We now make these observations quantitative by studying the probability $P(s; a,\phi)$ a given cluster has  $s$ particles. From  percolation theory we expect $P(s; a,\phi_c)$ to be gapped in  systems with a first-order transition, and to be gapless for a continuous transition \cite{saberi15,henkes16}.

\begin{figure}
	\centering
	\includegraphics[width=\columnwidth]{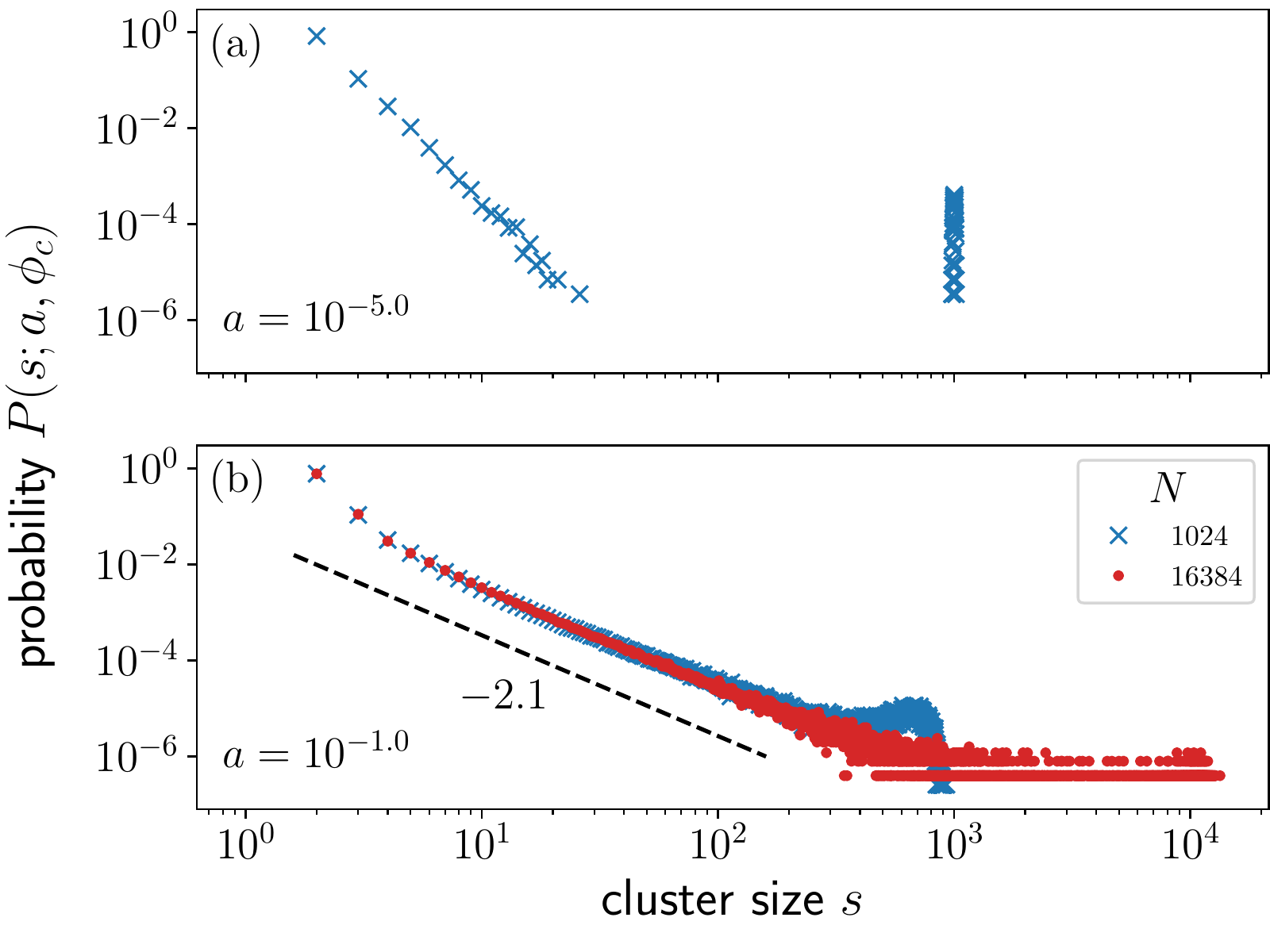}
	\caption{Cluster size probability distribution for (a) weakly attractive and (b) strongly attractive systems.}
	\label{fig:clusters}
\end{figure}

The cluster size distribution at $\phi_c$ is plotted in Fig.\,\ref{fig:clusters}a and b for weak and strong attraction, respectively. For weak attraction there is a clear gap between small clusters of tens of particles or less, and large clusters that contain nearly all particles in the packing, indicating a first order transition. We have verified that the large cluster peak is solely populated by jammed packings, while small clusters occur in both unjammed and jammed packings. 

The cluster size distribution for strongly attractive packings in Fig.\,\ref{fig:clusters}b shows no gap, indicating a continuous transition. We have verified that both jammed and unjammed packings populate the full range of cluster sizes. The distribution has a power law tail $P \sim s^{-\tau}$ that extends to cluster sizes of order $N$. To better estimate the exponent $\tau$, we plot the same distribution for a system of $N = 16384$ particles to find $\tau \approx 2.1$ (dashed line). The small peak for $s$ close to $N = 1024$ in the smaller systems is due to finite size effects, including the finite width of $f_j$. Note that the peak is reduced for larger $N$, while the distribution remains gapless.

\begin{figure}[tb]
	\centering
	\includegraphics[width=\columnwidth]{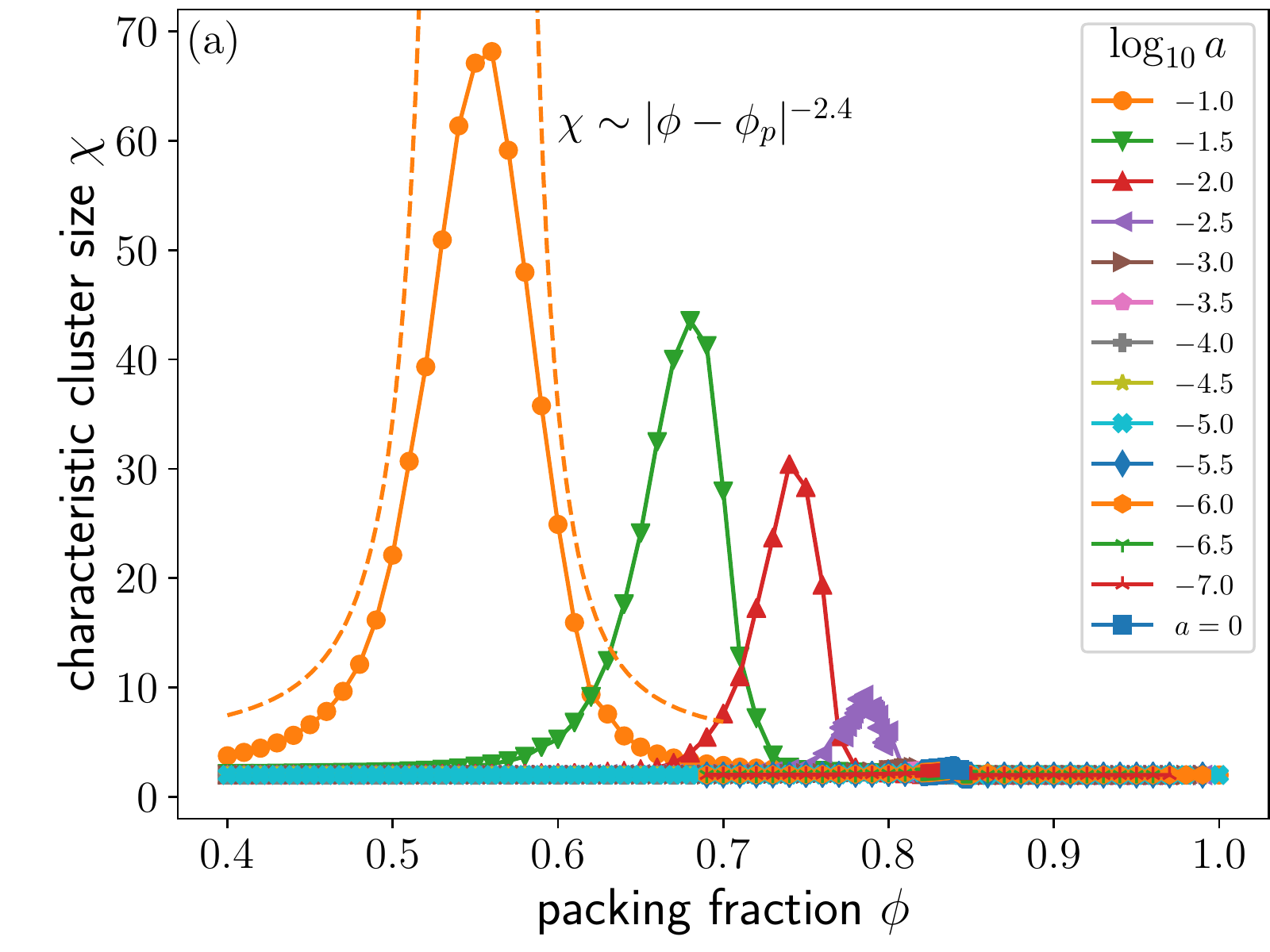}
	\includegraphics[width=\columnwidth]{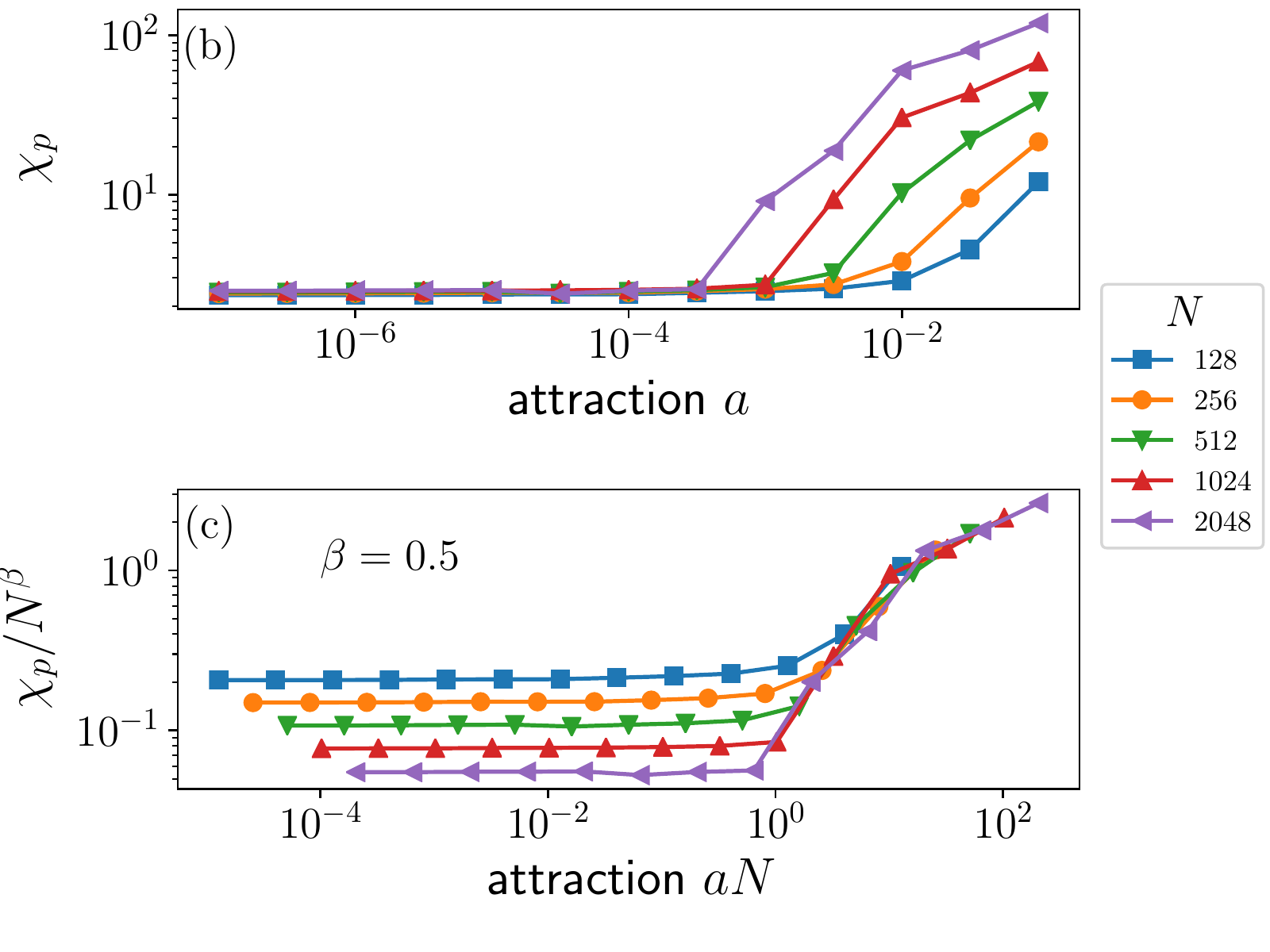}
	\caption{(a) Cluster size dependence on packing fraction. Dashed line shows inferred divergence of infinite system (offset vertically). (b) Evolution of the peak cluster size with attraction strength for varying particle number $N$. (c) Rescaled data from (b).}
	\label{fig:charbond}
\end{figure}

{\it Growing cluster size.}---
Having addressed statistics at $\phi_c$, we now probe cluster size as $\phi$ is swept through the jamming transition. Our results will further validate the first order and continuous characterization of weakly and strongly attractive jamming, respectively. Of equal importance, we will also identify the characteristic attraction strength $a^*$ separating weak and strong attraction.

For a continuous percolation transition, one expects to find a typical cluster size that diverges at the transition, while the same quantity should remain finite at a first order transition \cite{saberi15}. To quantify cluster sizes on either side of jamming, 
we introduce the probability $n(s; a,\phi)$ that a given non-spanning cluster has $s$ particles and calculate the expected cluster size of a randomly selected particle outside the spanning cluster,
\begin{equation}
\chi(a,\phi) = \frac{\sum_s s^2 \, n(s; a,\phi)}{\sum_s s \, n(s; a,\phi)} \,.
\end{equation}

In Fig.\,\ref{fig:charbond}a, $\chi$ is plotted versus packing fraction for varying attraction strength. While data for the lowest values of $a$ show no dramatic features, for the strongest attraction strengths there is a substantial increase in $\chi$ near $\phi_c$. To quantify these observations, we extract the height $\chi_p$ and position $\phi_p$ of the peak in $\chi$. From $\phi_p$ we calculate the shift $\phi_p^{\rm shift}(a) = \phi_c(0) - \phi_p(a)$. We find excellent agreement between the position of the peak and $\phi_c$ determined from Fig.~\ref{fig:fj}a, as demonstrated in the inset of Fig.~\ref{fig:fj}b. We conclude that the peak in $\chi$ coincides with the jamming point.

We now ask if the peak cluster size diverges as $N \rightarrow \infty$. Fig.\,\ref{fig:charbond}b shows $\chi_p$ as a function of $a$ for varying $N$. At low $a$, typical clusters consist of a few particles. There is no trend with $N$, suggesting that $\chi_p$ remains finite.
For strong attraction $\chi_p$ grows with $N$, and the attraction $a^*$ where $\chi_p$ starts to grow is lower in larger systems. To gain insight into these effects, in Fig.\,\ref{fig:charbond}c we replot the data as $\chi_p/N^\beta$ versus $aN$. We observe collapse to a master curve when $aN \gtrsim 1$ and $\beta \approx 0.5$. As $\beta$ is positive and the master curve increases with $aN$, we infer that $\chi_p$ diverges in the large system limit -- there is indeed a diverging cluster size, consistent with a continuous transition. For the largest $a$, the cluster size diverges as $\chi \sim 1/|\phi - \phi_p|^{2.4}$ (vertically offset dashed curve in Fig.~\ref{fig:charbond}a; log-log plot in the Supplementary Material). 

A key finding is that the rescaled attraction strength $aN$ in Fig.~\ref{fig:charbond}c implies the existence of a characteristic scale $a^* \sim 1/N$. Systems with $a$ above (below) $a^*$ jam according to the strongly (weakly) attractive scenario. Hence any nonzero attraction strength satisfies $a > a^*$ in a sufficiently large system, and in the $N \rightarrow \infty$ limit {\em all} attractive systems jam according to the strongly attractive scenario. In other words, attraction is never a weak perturbation to repulsive jamming. 
 
{\em Discussion.---} We have demonstrated that rigid clusters form a jammed phase in purely repulsive and weakly attractive systems via a first order transition in which the spanning cluster appears suddenly at the critical packing fraction. In sharp contrast, strongly attractive systems jam via a continuous transition with a typical cluster size that diverges at $\phi_c$. The first order transition for weak attraction is a finite size effect, and in thermodynamically large systems the jamming universality class is either purely repulsive ($a = 0$) or attractive ($a > 0$). As attraction is generic in experimental systems, we predict that they jam according to the attractive scenario.

Some of our results can be compared to work by Zheng et al.~\cite{zheng16} and Lois et al.~\cite{lois08}, with the caveat that preparation protocols differ. Zheng et al.~observed a critical packing fraction shift $\phi_c^{\rm shift} \sim a^{0.3}$, extracted from four values of $a$ over three decades; we find an exponent 0.5 with finer sampling. Lois et al.~\cite{lois08} report data for just one attraction strength comparable to our $a = 10^{-2}$. They found the fraction of jammed states collapses with $\alpha \approx 0.16$, and a cluster size  exponent  $\tau \approx 2.1$, in accord with our $\alpha \approx 0.2$ and $\tau \approx 2.1$. Henkes et al.~recently studied rigid clusters in frictional shear flow \cite{henkes16}. Despite the obvious differences between friction and attraction, they also found a continuous transition at nonzero friction.

There are several directions for future work. Foremost, it remains to determine the influence of rigid clusters on mechanics, such as storage and loss moduli \cite{tighe11,boschan16,baumgarten17b}, yield stress \cite{olsson07,rahbari10,gu14,irani14,irani16,tighe10c,dagois-bohy17}, nonlocal effects \cite{bocquet09,baumgarten17,tang18}, and shear banding \cite{singh14,irani14,irani16}. By varying the pair potential, once can also untangle the roles of the range and strength of the attractive interaction. The phase diagram for attractive glasses and gels has $\phi$ on one axis and the ratio of the attractive well depth $U$ to the thermal scale $k_B T$ on the other \cite{trappe01,trappe04}. Jammed states at $T = 0$ sit deep in the glass/gel phase, hence one anticipates connections to vitrification or gelation as $T$ increases.

We acknowledge financial support from the Nederlandse Organisatie voor Wetenschappelijk Onderzoek (Netherlands Organization for Scientific Research, NWO). This work was sponsored by NWO Exacte Wetenschappen (Physical Sciences) through the use of supercomputer facilities.

\end{document}